# Pseudopotential MRT lattice Boltzmann model for cavitation bubble collapse with high density ratio[*]

Ming-Lei Shan(单鸣雷)[1,2], Chang-Ping Zhu(朱昌平)[1,2,†], Cheng Yao(姚澄)[1], Cheng Yin(殷澄)[1], and Xiao-Yan Jiang(蒋小燕)[1]

[1]*College of Internet of Things Engineering, Hohai University, Changzhou 213022, China*

[2]*Jiangsu Key Laboratory of Power Transmission and Distribution Equipment Technology, Hohai University, Changzhou 213022, China*

The dynamics of the cavitation bubble collapse is a fundamental issue for the bubble collapse application and prevention. In present work, the modified forcing scheme for the pseudopotential multi-relaxation-time lattice Boltzmann model developed by Li Q. *et al.* is adopted to develop a cavitation bubble collapse model. In the respects of coexistence curves and Laplace law verification, the improved pseudopotential multi-relaxation-time lattice Boltzmann model is investigated. The independence between the kinematic viscosity and the thermodynamic consistency, surface tension is founded. By homogeneous and heterogeneous cavitation simulation, the capability of the present model to describe the cavitation bubble development as well as the cavitation inception is verified. The bubble collapse between two parallel walls is simulated. The dynamic process of collapsing bubble is consistent with the results from experiments and simulations by other numerical method. It is demonstrated that the present pseudopotential multi-relaxation-time lattice Boltzmann model is available and efficient, and the lattice Boltzmann method is an alternative tool for collapsing bubble modeling.

**Keywords:** lattice Boltzmann method, pseudopotential model, bubble collapse, improved forcing scheme

**PACS:** 47.11.Qr, 47.55.Ca, 47.55.dd, 47.55.dp

## 1. Introduction

Cavitation is a unique phenomenon in the liquid, and has attracted extensive research efforts both in science and engineering fields. Due to the various dynamics effects during the cavitation generation and cavitation bubble collapse, such as bubble wall oscillation and the formation of shockwave and micro-jet, the potential applications of cavitation has been and is being extensively explored in the fields of medicine, biology, lab on chip, sewage treatment, material surface cleaning treatment.[1-5]. However, as too many phenomena are involved, theoretical model of cavitation bubble collapse is difficult to be established, and for particular boundary conditions, the analytical solution is even impossible. Therefore, the numerical simulation becomes a powerful way to gain a better understanding. The conventional numerical simulation methods of cavitation bubble mainly include the finite volume method (FVM)[6-8], the finite element method (FEM)[9] and the boundary element method (BEM)[10]. In these partial differential equation based numerical simulations, the methods to track or capture the interfaces (such as the volume of fluid (VOF)[11] or level set method (LSM)[12]) are required, which are often extremely computationally demanding. In addition, Poisson equation needs to be solved to satisfy continuity equation, which drastically reduces the computational efficiency[13].

In recent years, the lattice Boltzmann method (LBM), which is a mesoscopic approach based on the kinetic Boltzmann equation, has emerged as a powerful tool for simulating multiphase flow problems[14-18]. As an alternative tool for the numerical simulations and investigations of multiphase flows, the LBM provides many advantages including clear physical pictures, easy implementation of boundary conditions, and fully parallel algorithms[19]. Generally, the existing LBM multiphase models can be mostly classed into four categories: the color-gradient method[20], the pseudopotential method (or Shan-Chen model)[21-22], the free-energy method[23-24], and the phase-field method[25]. The pseudopotential method, for its conceptual simplicity and computation efficiency, is widely and successfully used in LBM multiphase community. In pseudopotential method, the fluid interactions are mimicked by an interparticle potential, from which a non-monotomic equation of state (EOS) can be obtained. As a result, the separation of fluid phases or components can be achieved automatically in this method, and the methods to track or capture the interfaces are not required yet. Moreover, the pressure can be calculated from EOS efficiently instead of Poisson equation.

[*] Project supported by the National Natural Science Foundation of China (Grant Nos. 11274092 and 1140040119), the Natural Science Foundation of Jiangsu Province, China (Grant No. SBK2014043338)
[†] Corresponding author. E-mail: cpzhu5126081@163.com

Recently, the pseudopotential LBM was firstly introduced into the issue of cavitation by Sukop and Or[26]. In the following several years, research efforts have been made to investigate the mechanism of cavitation. Chen *et al.*[27] simulated the cavitating bubble growth using the modified pseudopotential LB model with the EDM force scheme. The results in quiescent flows agree fairly well with the solution of Rayleigh-Plesset equation. Mishra[28] introduced a model of cavitation based on the pseudopotential LB model that allows for coupling between the hydrodynamics of a collapsing cavity and supported solute chemical species. Using pseudopotential LB model, Daemi[29] modeled the bubble cluster in an acoustic filed, and verified that the deformation and coalescence phenomena in bubble cluster dynamics can be simulated by LBM. However, above mentioned research on cavitation does not include the cavitation bubble collapse process, which is essential for the study of cavitation phenomenon, and the density ratio between liquid phase and gas phase is limited under 100.

For pseudopotential LB model, the density ratio is a synthetic problem relating with lattice model, forcing scheme, thermodynamic consistency, numerical stability etc., and great efforts have been made for this issue[30-35]. In the present work, the modified forcing scheme for pseudopotential Multi-Relaxation-Time (MRT) LB developed by Li Q. *et al.*[33] is adopted to develop a cavitation bubble collapse model. Essentially, Li's scheme is an approximate approach for thermodynamic consistency by adjusting forcing scheme. This approach retains the concise and high efficiency of original LBM, and achieves the stable LB multiphase model with large density ratio.

The rest of the present paper is organized as follows. Section 2 will briefly introduce the pseudopotential MRT-LB model and the forcing scheme modified by Li *et al.* The numerical analysis of modified pseudopotential MRT-LB model for bubble will be given in Section 3. In Section 4, the validity and feasibility of the method for cavitation and bubble collapse will be verified. Finally, a brief conclusion will be made in Section 5.

## 2. Modified pseudopotential MRT-LB model

### 2.1. Pseudopotential MRT-LB model

The pseudopotential LB model, also known as Shan-Chen model, was developed by Shan and Chen in1993[21]. In pseudopotential method, the fluid interactions are mimicked by an interparticle potential, which is now widely called peseudopotential.

In original pseudopotential LB model, the Single-Relaxation-Time (SRT) collision operator was employed. In recent years, the MRT collision operator has been verified that it is superior to the SRT operator in terms of numerical stability. The MRT-LB evolution equation can be given as follows[33]:

$$f_\alpha(x+e_\alpha\delta_t, t+\delta_t) = f_\alpha(x,t) - (M^{-1}\Lambda M)_{\alpha\beta}(f_\beta - f_\beta^{eq}) + \delta_t F_\alpha', \tag{1}$$

where $f_\alpha$ is the density distribution function, $f_\beta^{eq}$ is its equilibrium distribution, $t$ is the time, $x$ is the spatial position, $e_\alpha$ is the discrete velocity along the $\alpha$ th direction, $\delta_t$ is the time step, $F_\alpha'$ is the forcing term in the velocity space, and $M$ is an orthogonal transformation matrix. $\Lambda$ in Eq.(1) is a diagonal matrix, and for D2Q9 lattice, $\Lambda$ given by

$$\Lambda = \mathrm{diag}(\tau_\rho^{-1}, \tau_e^{-1}, \tau_\zeta^{-1}, \tau_j^{-1}, \tau_q^{-1}, \tau_j^{-1}, \tau_q^{-1}, \tau_\upsilon^{-1}, \tau_\upsilon^{-1}). \tag{2}$$

Through the transformation matrix $M$, $f$ and $f^{eq}$ can be projected onto the moment space via $m = Mf$ and $m^{eq} = Mf^{eq}$, and the collision step of MRT-LB equation (1) can be rewritten as[33]

$$m = m - \Lambda(m - m^{eq}) + \delta_t\left(I - \frac{\Lambda}{2}\right)S, \tag{3}$$

where $I$ is the unit tensor, and $S$ is the forcing term in the moment space with $(I - 0.5\Lambda)S = MF'$. For D2Q9 lattice, $m^{eq}$ can be given by

$$m^{eq} = \rho\left(1, -2+3|v|^2, 1-3|v|^2, v_x, -v_x, v_y, -v_y, v_x^2 - v_y^2, v_x v_y\right)^T, \tag{4}$$

where $\rho = \sum_\alpha f_\alpha$ is the macroscopic density, $v$ is the macroscopic velocity and $|v|^2 = v_x^2 + v_y^2$. Then the streaming step of the MRT-LB equation can be formulated as

$$f_\alpha(x+e_\alpha\delta_t, t+\delta_t) = f_\alpha^*(x,t), \tag{5}$$

where $f^* = M^{-1}m^*$. The macroscopic velocity in Eq.(4) is calculated by

$$v = \left(\sum_\alpha e_\alpha f_\alpha + \frac{\delta_t}{2}F\right)\Big/\rho, \tag{6}$$

where $F = (F_x, F_y)$ for two dimensional space is the force action on the system.

For the pseudopotential LB model, the $F$ in Eq.(6) is given by[36]

$$F = -G\psi(x)\sum_{\alpha=1}^{N} w_\alpha \psi(x+e_\alpha)e_\alpha, \tag{7}$$

where $\psi(x)$ is the interparticle potential, $G$ is the interaction strength, and $w_{1-4} = 1/3$ and $w_{5-8} = 1/12$ are the weights. $F$ is incorporated via $S$ with specific forcing scheme. For MRT-LB method, the usual forcing scheme for D2Q9 lattice can be given by

$$S = \begin{bmatrix} 0 \\ 6(v_x F_x + v_y F_y) \\ -6(v_x F_x + v_y F_y) \\ F_x \\ -F_x \\ F_y \\ -F_y \\ 2(v_x F_x - v_y F_y) \\ (v_x F_y + v_y F_x) \end{bmatrix}. \quad (8)$$

### 2.2. Li's forcing scheme for pseudopotential MRT-LB model

For the pseudopotential multiphase LB model, the vapor density $\rho_g$, liquid density $\rho_l$ and the pressure $p_0$ in an equilibrium coexistence fluid, should satisfy the mechanical stability condition expressed as[33,36]:

$$\int_{\rho_g}^{\rho_l}(p_0 - p_{EOS-PL})\frac{\psi'}{\psi^{1+\varepsilon}}d\rho = 0, \quad (9)$$

where $\psi' = d\psi/\rho$, $\varepsilon$ is a parameter related with the interaction range[36]. $p_{EOS-PL}$ in the Eq.(9) is the non-ideal EOS of the pseudopotential LB model, which is given by

$$p_{EOS-PL} = \rho c_s^2 + \frac{Gc^2}{2}\psi^2, \quad (10)$$

where $c_s$ is the lattice sound speed. Using the Eq.(9) and (10), the coexistence curve of the pseudopotential LB model can be solved by numerical integration. From the thermodynamic theory view, the equilibrium coexistence fluid should also satisfy the Maxwell equal-area rule which determines the thermodynamic coexistence and is expressed as:

$$\int_{\rho_g}^{\rho_l}(p_0 - p_{EOS-TD})\frac{1}{\rho^2}d\rho = 0. \quad (11)$$

Here $p_{EOS-TD}$ is the EOS in the thermodynamic theory. Without special treatment, the mechanical stability condition decided by Eq.(9) will deviate the solution given by the Maxwell construction of Eq.(10). That is usually called the thermodynamic inconsistency in the pseudopotential LB model.

To address the thermodynamic inconsistency, the integrand in Eq.(9) and (11) should be equal or approximate to each other. Each integrand is naturally divided into two parts, i.e. the bracketed part which relates directly with EOS and the outside part which relates directly with $\rho$. So, two different strategies have been developed to solve the thermodynamic inconsistency. The first strategy is to guarantee the outside parts in Eq.(9) and (11) equal to each other at first[37], i.e. $\frac{\psi'}{\psi^{1+\varepsilon}} = \frac{1}{\rho^2}$. In this strategy framework, the specific form of $\psi$ is proposed as follows[37]

$$\psi(\rho) = \begin{cases} \exp(-1/\rho), & \varepsilon = 0 \\ \left(\dfrac{\rho}{\varepsilon+\rho}\right)^{1/\varepsilon}, & \varepsilon \neq 0 \end{cases}. \tag{12}$$

However, it is difficult for the first strategy to make $p_{EOS-PL} = p_{EOS-TD}$. In addition, the form EOS given by Eq.(10) is fixed. So, the adoption of different EOS is impossible.

The second strategy is to guarantee the bracketed parts in Eq.(9) and (11) equal to each other at first[38] ( i.e. $p_{EOS-PL} = p_{EOS-TD} = p_{EOS}$ ), then to reduce the deviation between $\dfrac{\psi'}{\psi^{1+\varepsilon}}$ and $\dfrac{1}{\rho^2}$. As a result, $\psi$ can be given by

$$\psi(\rho) = \sqrt{\dfrac{2(p_{EOS} - \rho c_s^2)}{Gc^2}}. \tag{13}$$

Here, $G$ is used to ensure that the whole term inside the square root is positive[38]. The second strategy allows different EOS to be involved, and is widely used in pseudopotential LB models[15].

In the second strategy framework, Li et al.[33] proposed a MRT version forcing scheme to achieve thermodynamic consistency. For the D2Q9 lattice, Li's forcing scheme can be given by

$$\mathbf{S} = \begin{bmatrix} 0 \\ 6(v_x F_x + v_y F_y) + \dfrac{0.75\varepsilon|\mathbf{F}|^2}{\psi^2 \delta_t (\tau_e - 0.5)} \\ -6(v_x F_x + v_y F_y) - \dfrac{0.75\varepsilon|\mathbf{F}|^2}{\psi^2 \delta_t (\tau_\varsigma - 0.5)} \\ F_x \\ -F_x \\ F_y \\ -F_y \\ 2(v_x F_x - v_y F_y) \\ (v_x F_y + v_y F_x) \end{bmatrix}, \tag{14}$$

Comparing with Eq.(8), $S_1$ and $S_2$ are modified with a term which can be tuned by $\varepsilon$. Through theoretical analysis, Li et al. founded that there exists a suitable $\varepsilon$ which can make the mechanical stability solution approximately identical to the solution given by the thermodynamic consistency requirement in a wide range of temperature[32-33]. Li's forcing scheme has the following advantages: (a) maintaining a uniform layout with the general form of the LB forcing scheme; (b) achieving thermodynamic consistency only by tuning one constant parameter; and (c) fully retaining the LBM's advantages of simple and efficient.

## 3. Numerical simulations and analyses

In this section, numerical investigations for bubble will be conducted with Li's improved pseudopotential LB model. Firstly, the improved model will be investigated for bubble simulations from three aspects: the optimal $\varepsilon$ for bubble simulations, the effect on coexistence curves from relaxation time $\tau_\upsilon$ and the validation of Laplace law. Then, the homogeneous and heterogeneous cavitation will be simulated to investigate the feasibility of presented LB model for cavitation phenomenon.

### 3.1. Investigation of improved model

Three numerical simulations of bubble are considered to investigate Li's improve pseudopotential MRT-LB model. The first is the simulation of stationary bubble, which can be used to obtain a numerical coexistence curve approximate to that given by the Maxwell construction by tuning $\varepsilon$. The second numerical simulation is to investigate the effect of relaxation time. The last simulation is conducted to validate the law of Laplace. In the present work, the Carnahan-Starling (CS) EOS is adopted, which can be given by[38]

$$p_{EOS} = \rho RT \frac{1+b\rho/4+(b\rho/4)^2-(b\rho/4)^3}{(1-b\rho/4)^3} - a\rho^2, \qquad (15)$$

where $a = 0.4963 R^2 T_c^2 / p_c$ and $b = 0.18727 RT_c / p_c$. Here $T_c$ and $p_c$ are the critical temperature and pressure, respectively. In our simulations, we set $a = 0.5$, $b = 1$, $R = 1$, $\delta_t = 1$. A 201×201 lattice is adopt in the simulations of this section. A bubble with radius of $r_0 = 70$ is initially placed at the center of the domain. The density field is initialized as[31]

$$\rho(x,y) = \frac{\rho_l + \rho_g}{2} + \frac{\rho_l - \rho_g}{2} \tanh\left[\frac{2\left(\sqrt{(x-x_0)^2+(y-y_0)^2} - r_0\right)}{W}\right], \qquad (16)$$

where $(x_0, y_0)$ is the center of the domain, $W$ is the prescribed width of the phase interface and is set as 5 in present work, 'tanh' is the hyperbolic tangent function and $\tanh(x) = (e^{2x}-1)/(e^{2x}+1)$. The periodical boundary conditions are applied in the $x$ and $y$ directions. Due to the use of CS EOS, $\psi$ adopts the form in Eq.(13), and $G = -1$ is used. The relaxation times are chosen as follows: $\tau_\rho^{-1} = \tau_j^{-1} = 1.0$, $\tau_e^{-1} = \tau_\zeta^{-1} = 1.1$, and $\tau_q^{-1} = 1.1$.

Li et al. have estimated that optimal value of $\varepsilon$ should be between 1 and 2[32]. For the stationary liquid droplet with a radius of $r_0 = 50$, they considered that $\varepsilon = 1.76$ is the optimal value. With this value, the coexistence curves obtained from LBM simulations are in good agreement with those given by the Maxwell equal-area construction. For stationary vapor bubble, the coexistence curves of the case $\tau_\upsilon = 0.6$ with different $\varepsilon$ are shown in Fig.1. From the figure we can see that, when $\varepsilon = 1.0$ the vapor branch of the coexistence curve significantly deviates from the solution of the Maxwell equal-area construction and the achievable lowest reduced temperature is around $0.7T_c$. For $\varepsilon = 2.0$, the LB model works well at $0.5T_c$, but there are still obvious deviations for the coexistence curves. In contrast, the coexistence curves as $\varepsilon = 1.86$ agrees quite well with the solution of the Maxwell equal-area construction in a wide range of temperature. The maximum density ratio is beyond 700, which is close to the density ratio of water and vapor in the nature. Fig.1 demonstrates that, for vapor bubble, the improved MRT pseduopotential LB model is capable of achieving thermodynamic consistency, large density ratio and stability by tuning $\varepsilon$. In addition, it is should be noted that the optimal value of $\varepsilon$ for bubble case is larger than that for droplet case. This may be associated with the effect of surface tension and the difference of compressibility between vapor phase and liquid phase.

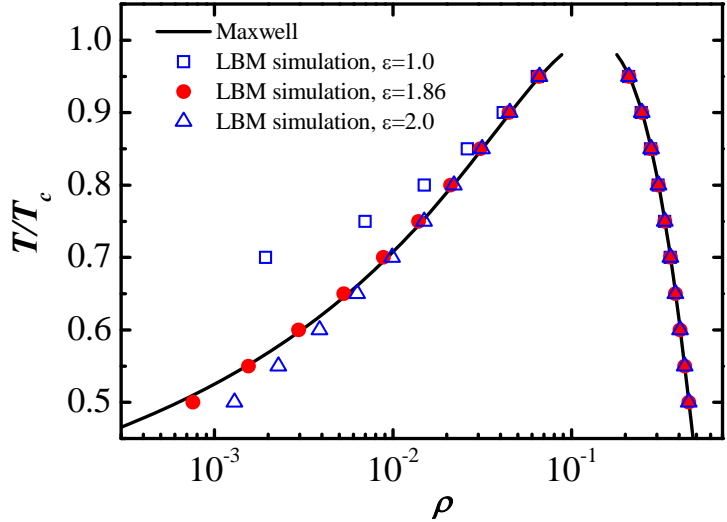

Fig. 1. (Color online) Comparison of the numerical coexistence curves by LBM with different $\varepsilon$ ($\tau_\upsilon = 0.6$)

Furthermore, the effect of relaxation time $\tau_\upsilon$, which is associated with kinematic viscosity, is investigated. The coexistence curves of the cases $\tau_\upsilon = 0.51$, $\tau_\upsilon = 0.6$ and $\tau_\upsilon = 0.8$ as $\varepsilon = 1.86$ are shown in Fig.2. We can see that good agreement can also be achieved. It demonstrates that thermodynamic consistency is independent of kinematic viscosity for Li's improved MRT pseduopotential LB model.

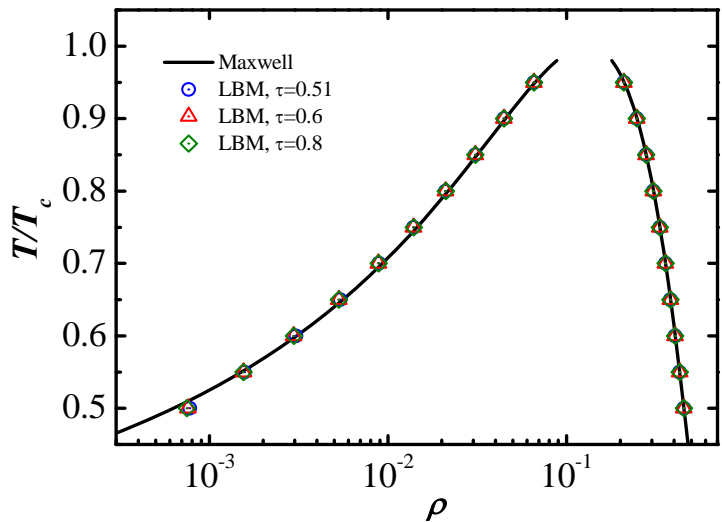

Fig. 2. (Color online) Numerical coexistence curves by LBM with different $\tau_\upsilon$ ($\varepsilon = 1.86$)

The satisfaction of Laplace law is an important benchmark test for multiphase flows. The surface tension force, meanwhile, can be calculated when Laplace law is verified. For the bubble case, the Laplace law can be given as

$$\Delta p = p_{out} - p_{in} = \frac{\sigma}{r}, \tag{17}$$

where $p_{out}$ and $p_{in}$ are the fluid pressures outside and inside of the bubble, respectively. $\sigma$ is the surface tension and $r$ is the radius of the bubble. The results at different reduced temperatures are shown in Fig.3 for the cases $\tau_\upsilon = 0.6$ and $\tau_\upsilon = 0.8$. From Fig.3, the linear relationship with $\Delta p$ and $1/r$ can be clearly observed. The slop of the lines, $\sigma$, are almost same for $\tau_\upsilon = 0.6$ and $\tau_\upsilon = 0.8$, and decrease as temperature increasing. It indicates that the improved MRT pseduopotential LB model satisfies Laplace law. And the independence of $\sigma$ from $\tau_\upsilon$ is more convenient for the physical mechanism investigation of the multiphase flows.

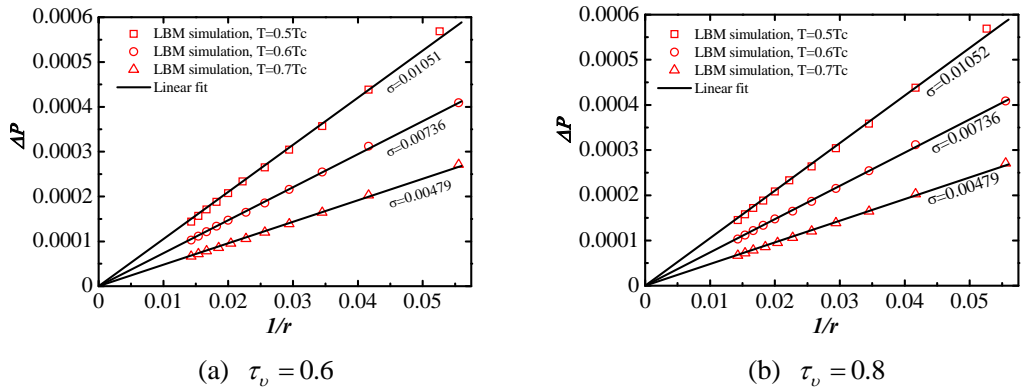

(a) $\tau_\upsilon = 0.6$     (b) $\tau_\upsilon = 0.8$

Fig. 3. (Color online) Numerical coexistence curves by LBM with different $\tau_\upsilon$ ($\varepsilon = 1.86$)

### 3.2. Cavitation inception and development

The inception of cavitation bubble is essentially a phase transition from liquid to vapor. It happens when the pure liquid is exposed to such a high negative pressure that makes the density of liquid being in the range of coexistence density.

In present work, the cavitation inception phenomenon is investigated by proposed MRT pseduopotential LB model. A $(lx,ly) = (201,201)$ lattice is adopted. The whole density field is initialized as $\rho_{init} = 0.2609$ which is between the critical density $\rho_c = 0.1305$ and the liquid density $\rho_l = 0.4541$ at $0.5T_c$ with $\tau_\upsilon = 0.6$ and $\varepsilon = 1.86$. Then a density perturbation of $\rho_{init}/10^8$ is set at $(ly-1)/2$. The periodical boundary conditions are applied in $x$ and $y$ directions. The densities at $((lx-1)/2, ly-1)$ and $((lx-1)/2,(ly-1)/2)$ are detected. The densities evolutions with time steps are shown in Fig. 4. In our simulation, the cavitation occurs as a linear band because of the initialed density perturbation is a line and the boundary conditions are periodical. At the first stage, the liquid phase and the vapor phase are not separated apparently. At about 180 time steps, the vapor phase is firstly separated at $(ly-1)/2$. After severe density fluctuations, the density at $((lx-1)/2,(ly-1)/2)$ is stable near the density of vapor phase $\rho_g = 6.269 \times 10^{-4}$. In contrast, the density at $((lx-1)/2, ly-1)$ has fluctuated for much more time before closing to the density of liquid phase. At last, a stable fluid system including liquid phase and vapor phase is formed. This simulation demonstrated that the proposed MRT pseduopotential LB model is available to describe the process of cavitation inception essentially.

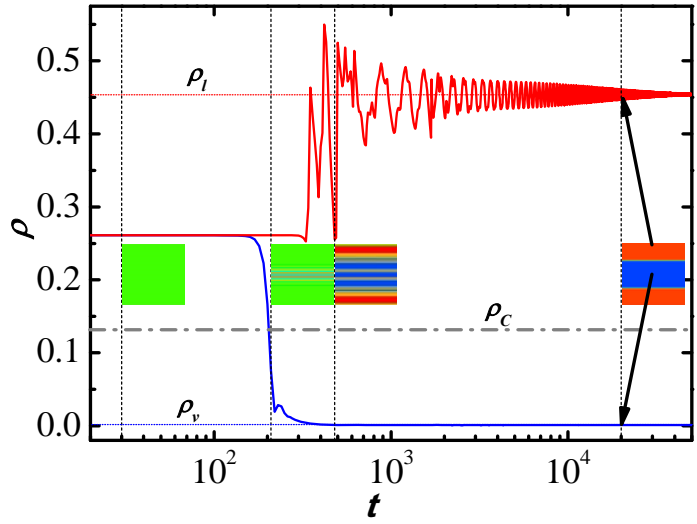

Fig. 4. (Color online) Density evolution during the process of cavitation inception simulated by LBM

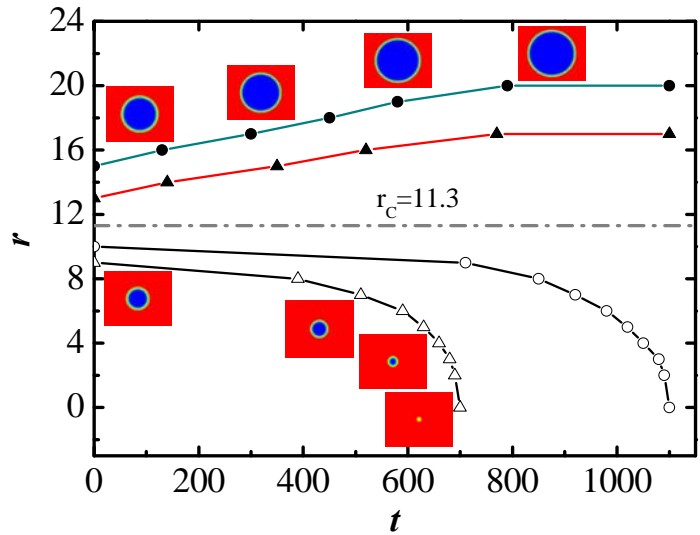

Fig. 5. (Color online) Effect of the cavitation nuclei critical radius verified by LBM

If there is a cavitation nucleus in the static fluid, the later development of the cavitation bubble is decided by the critical radius of cavitation nucleus, which can be given as follows[39]

$$r^* = -\frac{\sigma}{\Delta P} \quad (18)$$

where $r^*$ is the critical radius of the cavitation bubble, $\sigma$ is the surface tension and $\Delta P$ is the pressure difference between vapor and liquid phases relative to the flat interface. $\sigma$ can be get from the Laplace law verification in section 3.1, and the surface tension at $0.5T_c$ is used here, i.e. $\sigma = 0.01051$. Specific $\Delta P$ can be achieved by pressure boundaries[26] or by setting the density of liquid domain artificially. The second approach is used here. By tuning the liquid phase, a critical radius of 11.3 is achieved finally. In this simulation, a $401 \times 401$ lattice is adopted and the periodical boundary conditions are applied in $x$ and $y$ directions. For verify the effect of cavitation nuclei critical radius, the cavitation bubbles with radii greater and smaller than critical radius are initialized at the central of simulation domain, respectively. The evolutions of the bubbles are shown in Fig.5. We can see that a bubble with radius just below the critical value cannot overcome the energy barrier and eventually condenses. While, a bubble with radius just above the critical value expands gradually and reaches a new equilibrium eventually. It demonstrates that the proposed MRT pseduopotential LB model is valid to simulate the cavitation bubble development as well as the cavitation inception.

## 4. Bubble collapse between two parallel solid walls

A set of investigations for the proposed MRT pseduopotential LB model demonstrate that the LB model can establish cavitation numerical models in line with the law of physics. And with the stability including such a high density ratio closing to water and vapor in real world, the LB model is promising to establish an efficient cavitation numerical model. In this section, we will establish a numerical model to investigate the mechanism of the bubble collapse between two parallel solid walls.

The dynamics of cavitation bubble near solid wall is an important issue for understanding the mechanism of surface damages in fluid machineries. It has been widely studied and most of these studies are based on the semi-infinite fluid domain[40-42]. However, in the case of the bubble between two parallel solid walls, the collapse of the bubble will be significantly different from those in the semi-infinite assumption[3,43]. Although several investigations have been conducted based on macroscopic methods, the LB modeling of cavitation bubble collapse between two solid walls is still in its infancy. In this section, using the proposed MRT pseduopotential LB model with high density ratio, we make an attempt to simulate bubble collapse between two parallel solid walls in 2D case.

### 4.1. Computational domain

The computational domain for the bubble collapse between two parallel solid walls is shown in Fig. 6. In present simulation, a 1801×301 lattice is adopted. Using Eq. 16, a spherical bubble with the radius of $R_0$ is initialized in a stationary fluid between two parallel walls. $H$ is the distance between upper and lower walls, i.e. $H = 301$. And $d$ is the distance between the bubble center and the lower wall. $P_v$ and $P_\infty$ represent the pressure of vapor inside the bubble and the ambient pressure outside the bubble.

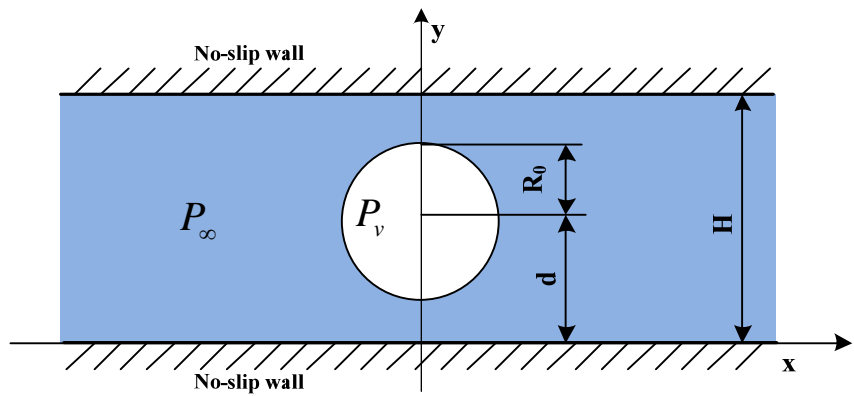

Fig. 6. (Color online) Schematic diagram of the computational domain

For describing the initial simulation states quantitatively, two non-dimensional parameters are introduced as follows

$$\lambda = \frac{R_0}{d}, \tag{19}$$

$$\kappa = \frac{d}{H}, \tag{20}$$

where $\lambda$ is defined as the stand-off distance between the bubble center and the lower wall, $\kappa$ is defined as the relative location of the bubble to the channel between two walls. In this simulation, $\kappa$ is set as 0.5, i.e. the bubble center locates at the horizontal centerline of the channel; $\lambda$ is as 0.92 according to Ref.[44]. So the parameters of computational domain can be assigned, i.e. $d=150$ and $R_0=138$.

In order to simulate the bubble collapse process, a positive pressure difference $\Delta P = P_\infty - P_v$ is achieved by artificially tuning the initial liquid density based on the equilibrium state. The periodical boundary condition is applied in $x$ direction, and no-slip bounce-back boundary condition is implemented in $y$ direction. The parameters of the LB model are set as $T/T_c = 0.5, \tau_\upsilon = 0.51$ and $\varepsilon = 1.86$.

### 4.2. Simulation results

The simulation results of bubble collapse between two parallel walls by the proposed MRT pseduopotential LB model are shown in Fig. 7 compared with the experimental results[44] and the VOF simulations[43]. In the whole collapse process, the longitudinal length of the bubble is almost fixed due to the retarding effect of two parallel walls. In contrast, the bubble laterally shrinks step by step to be dumbbell type, and then splits in two bubbles after the collapse. Finally, the jets towards upper and lower walls are formed. This dynamics process keeps consistent with the experiments and VOF simulations.

To investigate the dynamics characteristics in more details when the bubble collapses and the jet formats, the pressure distribution, the regional density and velocity distributions at these two key moments are shown in Fig. 8 and Fig. 9. The dumbbell type bubble splits into two sub-bubbles at the moment of collapse. From Fig. 8 we can find that the impetus of collapse comes from the horizontal liquid flows to the bubble neck. A colliding interface can be found obviously at the tips of two cone-shaped sub-bubbles. As a result, a higher pressure/density region is formatted at the bubble neck. This higher pressure/density region becomes the leading source power formatting the jets.

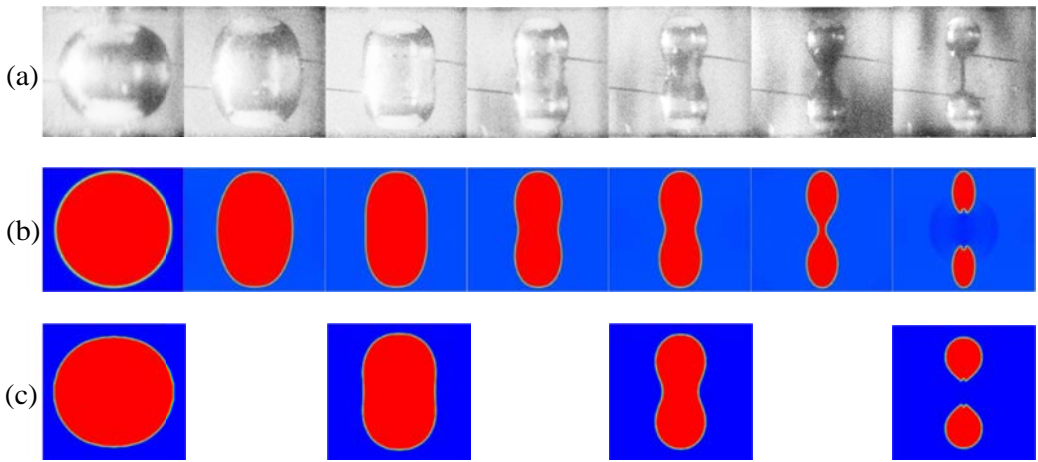

(a)

(b)

(c)

Fig. 7. (Color online) Comparison of results by: (a) experiments, (b) LBM simulations, (c) VOF simulations

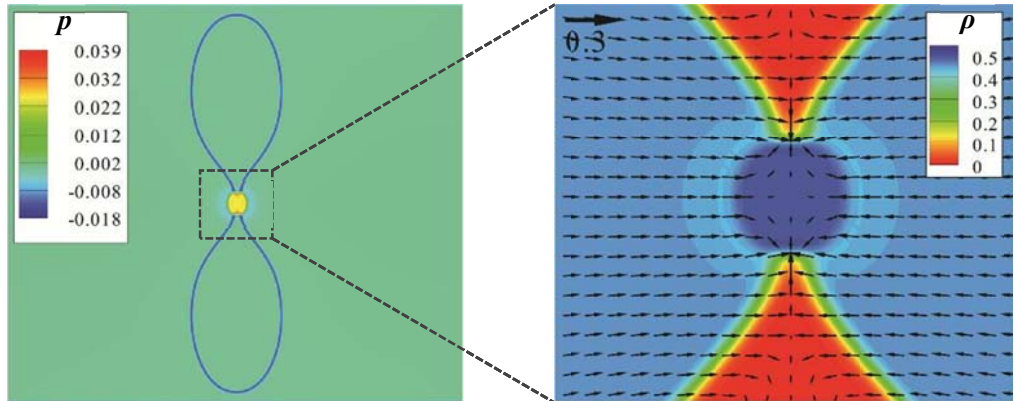

Fig. 8. (Color online) Pressure distribution (left) and the regional density and velocity distributions (right) as bubble collapse

Due to the regional higher pressure at the tips of sub-bubbles, the tips are flatted and then sag. Accordingly, there emerges the high flow velocity towards two sub-bubbles inner, respectively. Then the jets of collapsing bubble are formated, as shown in Fig. 9. The high velocity jet is considered as the leading factor to the wall damage. The detailed investigations of the interaction mechanisms between the high velocity jet of collapsing bubble and the parallel walls will be developed by present LBM model in our future work.

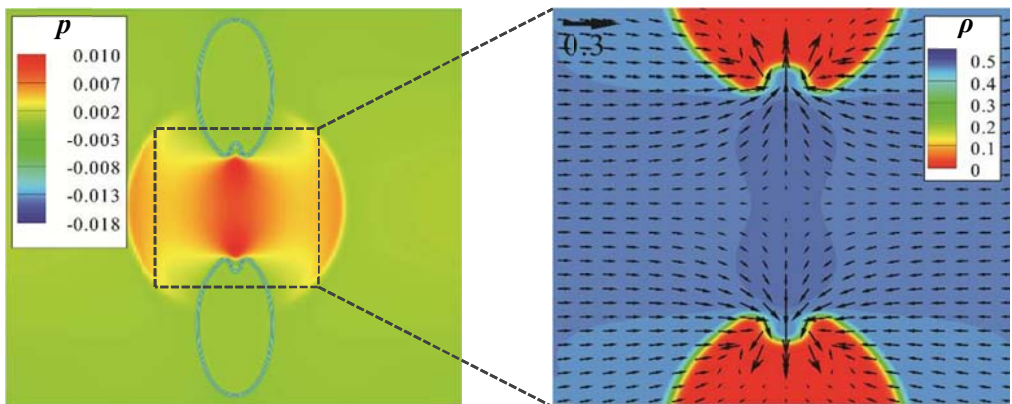

Fig. 9. (Color online) Pressure distribution (left) and the regional density and velocity distributions (right) as jet formatted

## 5. Conclusion

For the modeling of collapsing bubble, an improved MRT pseduopotential LB model was investigated in some respects such as thermodynamic consistency, Laplace law, homogeneous and heterogeneous cavitation. Then the bubble collapse between two parallel walls was simulated. The dynamic process of collapsing bubble is consistent with the results from experiments and simulations by other numerical method.

The improved forcing scheme developed by Li Q. *et al*. provides a convenient and efficient approach to achieve thermodynamic consistency. Adopting Li's forcing scheme, we found that the thermodynamic consistency and surface tension are independent of kinematic viscosity, which makes it superior to investigate the physical mechanism of the multiphase flows. Even at $0.5T_c$ with $\tau_\upsilon = 0.51$, the improved MRT pseduopotential LB model also has enough stability to describe the collapsing bubble with a high density ratio beyond 700. It is demonstrated that the present MRT pseduopotential LB model is available and efficient, and the LBM is an alternative tool for collapsing bubble modeling.